\documentclass[sigconf]{acmart}

\usepackage{comment}
\usepackage{tcolorbox}
\usepackage{listings}

\colorlet{punct}{red!60!black}
\definecolor{background}{HTML}{EEEEEE}
\definecolor{delim}{RGB}{20,105,176}
\colorlet{numb}{magenta!60!black}

\lstdefinelanguage{json}{
    basicstyle=\small\ttfamily,
    numbers=left,
    numberstyle=\scriptsize,
    stepnumber=1,
    numbersep=8pt,
    showstringspaces=false,
    breaklines=true,
    frame=lines,
    backgroundcolor=\color{background},
    literate=
     *{0}{{{\color{numb}0}}}{1}
      {1}{{{\color{numb}1}}}{1}
      {2}{{{\color{numb}2}}}{1}
      {3}{{{\color{numb}3}}}{1}
      {4}{{{\color{numb}4}}}{1}
      {5}{{{\color{numb}5}}}{1}
      {6}{{{\color{numb}6}}}{1}
      {7}{{{\color{numb}7}}}{1}
      {8}{{{\color{numb}8}}}{1}
      {9}{{{\color{numb}9}}}{1}
      {:}{{{\color{punct}{:}}}}{1}
      {,}{{{\color{punct}{,}}}}{1}
      {\{}{{{\color{delim}{\{}}}}{1}
      {\}}{{{\color{delim}{\}}}}}{1}
      {[}{{{\color{delim}{[}}}}{1}
      {]}{{{\color{delim}{]}}}}{1},
}

\AtBeginDocument{%
  }

\begin{document}

\title{MalCVE: Malware Detection and CVE Association
Using Large Language Models}

\author{Eduard Andrei Cristea}
\affiliation{%
  \institution{Norwegian University of Science and Technology}
  \city{Trondheim}
  \country{Norway}}
\email{falniir@gmail.com}

\author{Petter Molnes}
\affiliation{%
  \institution{Norwegian University of Science and Technology}
  \city{Trondheim}
  \country{Norway}
}
\email{petter@molnes.dev}

\author{Jingyue Li}
\affiliation{%
 \institution{Norwegian University of Science and Technology}
 \city{Trondheim}
 \country{Norway}}
\email{jingyue.li@ntnu.no}

\renewcommand{\shortauthors}{Cristea, Molnes, and Li}

\begin{abstract}
Malicious software attacks are having an increasingly significant economic impact. Commercial malware detection software can be costly, and tools that attribute malware to the specific software vulnerabilities it exploits are largely lacking. Understanding the connection between malware and the vulnerabilities it targets is crucial for analyzing past threats and proactively defending against current ones. In this study, we propose an approach that leverages large language models (LLMs) to detect binary malware, specifically within JAR files, and uses LLM capabilities combined with retrieval-augmented generation (RAG) to identify Common Vulnerabilities and Exposures (CVEs) that malware may exploit. We developed a proof-of-concept tool, MalCVE, that integrates binary code decompilation, deobfuscation, LLM-based code summarization, semantic similarity search, and LLM-based CVE classification. We evaluated MalCVE using a benchmark dataset of 3,839 JAR executables. MalCVE achieved a mean malware-detection accuracy of 97\%, at a fraction of the cost of commercial solutions. In particular, the results demonstrate that \textbf {\textit{LLM‑based code summarization enables highly accurate and explainable malware identification.}} MalCVE is also the first tool to associate CVEs with binary malware, achieving a recall@10 of 65\%, which is comparable to studies that perform similar analyses on source code.
\end{abstract}



\maketitle

\section{Introduction}
Global ransomware costs are projected to reach 276 billion USD annually by 2035 \cite{ransomware-statistics}. These figures do not account for other types of malware, indicating that the total economic impact of all malware is even greater.

The field of malware detection has made significant strides. However, a critical gap remains: attributing malware to the specific software vulnerabilities it exploits \cite{joyce2023maldictbenchmarkdatasetsmalware}. Understanding this relationship is essential for proactively defending against potential threats.
In 2017, the WannaCry ransomware spread across computers running the Windows operating system (OS) by exploiting the CVE-2017-0144 vulnerability. Nearly two months before the attack, Microsoft had released a patch for this vulnerability. However, many organizations had not applied the patch \cite{kaspersky-wannacry}.

Existing malware detection in the industry relies on proprietary and expensive tools (e.g., VirusTotal \cite{virustotal-pricing} and CrowdStrike \cite{crowdstrikefalcon:online}), which often lack an explainable connection to vulnerabilities. Surveys on explainable AI in cybersecurity show that many commercial malware detectors lack transparency, making it difficult to trace alerts back to specific vulnerabilities or code patterns \cite{Thesurve48:online}.
State-of-the-art research in malware detection and vulnerability classification employs fine-tuned models to detect malware, classify its type, or identify vulnerabilities in source code snippets \cite{zhou2024largelanguagemodelvulnerability}, \cite{malbert}. However, these fine-tuning approaches do not scale to vulnerability detection in obfuscated binaries, primarily because the field lacks a large-scale dataset for this purpose.
To our knowledge, no existing solution offers transparent, low-cost, and explainable malware-to-vulnerability attribution directly from obfuscated malicious binaries, especially without relying on domain-specific training.

Recent advances in LLMs and RAG offer a promising opportunity in this field \cite{nvidia-rag}, \cite{lewis2021retrievalaugmentedgenerationknowledgeintensivenlp}. The latest LLM models can process between 100,000 and 1,000,000 tokens, where each token corresponds to approximately 4 characters. This capability enables LLMs to analyze extensive codebases without losing contextual information \cite{google-malware-analysis}.
This study investigates whether LLMs, in combination with RAG, can effectively detect malware and identify the specific vulnerabilities it targets directly from decompiled Java binaries. The research is guided by the following questions.

\begin{itemize}
     \item \textbf{RQ1:} Can LLMs accurately classify decompiled binaries as either malicious or benign?
     \item \textbf{RQ2:} Can LLMs effectively and accurately associate malicious binaries with the specific CVEs the binaries exploit, using prompting and RAG techniques?
\end{itemize}

To answer the research questions, we propose approaches that incorporate static decompilation and deobfuscation, LLM-based code summarization, semantic similarity search for relevant vulnerabilities, and LLM-based vulnerability classification. We implemented these approaches in a prototype tool called MalCVE. Its performance was evaluated using accuracy metrics for both RQ1 and RQ2 on the MalDICT dataset \cite{joyce2023maldictbenchmarkdatasetsmalware}, which contains 3,839 JAR executables. Furthermore, RQ2 is measured with recall@k. Results show that MalCVE can detect if a binary software is malicious or benign, achieving an average accuracy of 97\%. The cost of detecting malware per file is low compared to commercial malware detectors, approximately 1/66 of the cost of CrowdStrike Falcon \footnote{https://www.crowdstrike.com/en-us/} and 1/80 of the cost of ANY.RUN \footnote{https://any.run/}. The tool can also list and rank the top 10 CVEs associated with the detected malware.  
The contributions of this study are:
\begin{itemize}
    \item We conducted a novel study providing comprehensive empirical evidence that generic LLMs are capable of detecting malware and identifying the specific vulnerabilities the malware exploits.
    \item Most components used to implement the approaches in the proof-of-concept tool MalCVE are open source, except for the LLMs, which may be commercial models. This design lowers the barrier to entry for individual researchers, academic institutions, and small enterprises to perform malware analysis, not only by reducing costs but also by minimizing the need for specialized expertise.
\end{itemize}

The remainder of this paper is organized as follows. Section \ref{sec:related-work} introduces related work. Section \ref{sec:system-design} presents the proposed approach. Section \ref{sec:evaluation} details the evaluation design and presents the results. Section \ref{sec:discussion} discusses the results. Section \ref{sec:conclusion} concludes the study and outlines future research.

\section{Related Work}
\label{sec:related-work}
This section reviews recent studies that utilize LLMs for the security analysis of both malware and source code.

\subsection{Malware Detection Using LLMs}

\citet{malbert} proposed MalBERT, a malware detection and categorisation solution for Android APKs. Their approach fine-tunes a pre-trained model on 12,000 benign applications and 10,000 malicious applications. The final solution achieved 97.6\% accuracy for binary classification and 91\% for malware type classification. 

\citet{google-malware-analysis} presents an in-depth test of Gemini 1.5 Pro, highlighting its application in malware detection. The study also presents examples that showcase the model's proficiency in identifying malicious code and summarizing its behavior and key functionality. The model successfully detected a zero-day attack that antivirus software had not previously flagged. This demonstrates that LLMs have capabilities that surpass those of traditional methods, such as antivirus software. \citet{scaling-malware-analysis-gemini-1-5-flash} provides an example demonstrating LLM’s capability to identify false positives. A file flagged by an antivirus vendor on VirusTotal \cite{virustotal-pricing} was correctly identified as benign by the LLM. A key limitation of the paper \cite{scaling-malware-analysis-gemini-1-5-flash} was that it presented only a few examples without systematic quantitative evaluations. 

Li et al. \cite{PE-Ransomware-Image-Detection} developed a novel ransomware-detection methodology that uses open-source LLMs and image analysis of portable executable (PE) files. Their research focused on four types of ransomware and four types of benignware. The results demonstrate the capability of LLMs to distinguish malware from benignware, with false-positive and false-negative rates below 7\% across the four types analyzed. Although the results of this research are promising, the study comprises only 221 files across eight file types. The dataset is small and imbalanced, which could skew the results.
\citet{walton2025exploringlargelanguagemodels} used OpenAI's GPT-4o-mini model to generate summaries for code at the function, class, and package levels for Android APK malware. Using these summaries, the model attempts to determine whether the Android APKs are malicious or benign. Using prompt engineering alone, their research achieved \textbf{77\%} classification accuracy. However, the dataset consists of only 200 files. Additionally, the research does not address code obfuscation techniques \cite{code-obs}. Code obfuscation is the process of transforming code into a functionally equivalent but difficult-to-understand form, a common approach for attackers to confuse malware detectors and cause malicious files to be classified as benign. 

\subsection{Vulnerability Detection in Source Code Using LLM and RAG}

\citet{vulrag}  found that current machine learning-based solutions struggle to achieve a deep understanding of vulnerability semantics, particularly when distinguishing between subtly different but semantically distinct code pairs. Their solution addresses this problem by integrating RAG. A tool named  
Vul-RAG was developed to demonstrate the idea, achieving a 61\% accuracy on the benchmark dataset they created. Furthermore, their user study showed that LLM-generated knowledge increased manual vulnerability detection by 17\%, confirming its value.
Although Vul-RAG outperformed state-of-the-art solutions, false negatives often occurred due to missing or inaccurate vulnerability knowledge, i.e., insufficient or incorrect context for the LLM.

\citet{LLM-CloudSec} introduced LLM-CloudSec, an unsupervised framework that employs LLMs for fine-grained vulnerability detection and analysis in cloud applications. Their aim is to overcome the limitations of traditional static and dynamic analysis and of existing machine learning-based methods by incorporating RAG and the Common Weakness Enumeration (CWE) as external knowledge bases. LLM-CloudSec comprises two collaborating LLM agents: a detection agent and an analysis agent. The detection agent uses few-shot learning with embedded CWE category names and numbers to identify whether a given code snippet contains a vulnerability and to classify it into one or more CWE categories via RAG. Upon a positive detection, the analysis agent retrieves a detailed description of the CWE, performs line-level localization, and generates an in-depth natural-language analysis of the vulnerability. Both agents rely on GPT-4-0613 as the underlying LLM. LLM-CloudSec was tested on a dataset comprising 10 instances of vulnerable code across 60 distinct CWEs from the Juliet C++ dataset. The detection agent achieved 70\% accuracy in classifying CWE vulnerabilities.

\section{Our approach and MalCVE implementation}
\label{sec:system-design}
Although existing studies have explored the potential of LLMs for identifying binary malware, they are limited to evaluations on small or imbalanced datasets and have not investigated the impact of code obfuscation. In terms of associating malware with vulnerabilities, existing studies are limited to source code analysis and CWE. While identifying CWEs is valuable for addressing the root causes of vulnerabilities, identifying CVEs is more valuable, as it provides actionable insights for remediation. Our approach aims to address these limitations. This section outlines the design principles of our approach and describes the implementation of the proof-of-concept tool, MalCVE.

\subsection{Approach design}

Our approach consists of eight steps to analyze a JAR and answer \textbf{RQ1} and \textbf{RQ2}. The steps are as follows: 
\begin{enumerate}
    \item Decompiling the JAR file
    \item Deobfuscating the code
    \item Summarizing the code using LLM
    \item Generating search queries
    \item Performing similarity search in the database
    \item Ranking the results
    \item Associating malicious JAR files with relevant CVEs
    \item Saving the results to a file
\end{enumerate}

\subsubsection{Decompilation}
\label{method:decompilation}

The first step in our process is decompiling the JAR executable, which is a prerequisite for the subsequent summarization step. The most widely adopted commercial disassembler and decompiler is IDA Pro \footnote{https://hex-rays.com/ida-pro}, which supports a wide array of architectures and file formats. IDA Pro is widely regarded as the industry standard for software reverse engineering \cite{IDAPropl44:online}. It employs signature matching FLIRT \footnote{https://docs.hex-rays.com/user-guide/signatures/flirt}, which recognizes standard library functions generated by supported compilers \cite{FLIRTHex92:online} and Lumina \footnote{https://hex-rays.com/lumina}, which tracks metadata about some widely recognizable functions, such as their names, prototypes, or operand types \cite{lumina86:online}.
Binary Ninja \footnote{https://binary.ninja/} provides a simplified GUI and a well-defined intermediate language, facilitating both manual and automated analysis. Its built-in decompiler and multi-language SDK (Python, C++, Rust) make it particularly popular among vulnerability researchers seeking to script complex analyses. While IDA Pro is considered state-of-the-art, several specialized open-source alternatives are available for specific programming languages. CFR \footnote{https://www.benf.org/other/cfr/} is an open-source decompiler specialized for Java. CFR supports the decompilation of modern Java features, yet it is implemented in Java 6, allowing compatibility with older Java files. In addition, CFR can convert other JVM languages, such as Kotlin or Scala, back into Java \cite{cfr-decompiler}. According to Strobel \cite{procyon-decompiler}, CFR is a better open-source decompiler for obfuscated JARs. Procyon \footnote{https://github.com/ststeiger/procyon} is another open-source decompiler for Java \cite{procyon-decompiler}. This decompiler also excels at newer Java features.
Ghidra \footnote{https://ghidra-sre.org/}, released by the NSA in 2019, has rapidly gained popularity in the open-source community. Its decompiler supports a wide range of architectures and translates binaries into human-readable pseudo C/C++ code.

We employ a combination of two decompilers—CFR 0.152 and Procyon 0.6.0—to leverage open-source solutions and minimize operational costs. Using two distinct decompilers provides redundancy, ensuring that if one fails to decompile a given JAR file, the other may succeed. Ghidra was also tested for this use case, as it is a popular open-source choice. However, CFR and Procyon are better suited to JAR files than Ghidra, as they generate actual Java source code rather than pseudocode.
MalCVE attempts to use the CFR decompiler as the primary option. If CFR fails to decompile the JAR, the system falls back to Procyon. In most cases, at least one of the decompilers successfully processes the file. If both fail, the file is excluded from further analysis. 

\subsubsection{Deobfuscation} \label{method:deobfuscation}

In the second step, a deobfuscation tool was applied. Deobfuscating the code ensures that the LLM can sufficiently understand the program's behavior, generate accurate search queries based on the libraries, methods, and classes used, and determine the nature of the code.

\begin{figure*}
    \centering
    \includegraphics[width=1\linewidth]{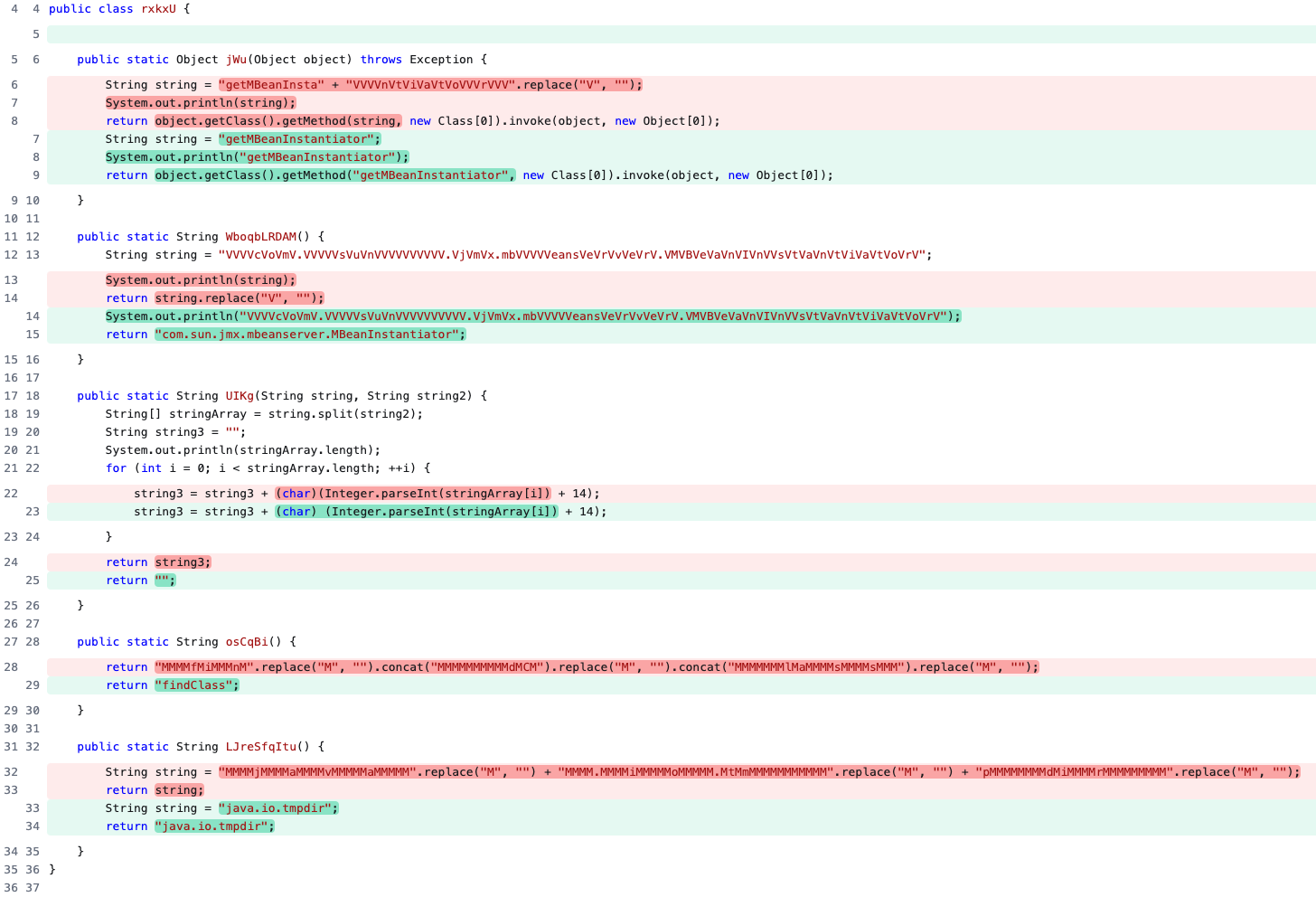}
    \caption{Before and After Decompilation Comparison}
    \label{fig:code-diff-deobfuscated}
\end{figure*}

Java deobfuscation tools aim to reverse common obfuscation techniques used to hinder reverse engineering. Several general-purpose tools are available for this purpose, such as Java Deobfuscator \footnote{https://github.com/java-deobfuscator/deobfuscator}. Many of these deobfuscators have limitations, such as outdated codebases, lack of maintenance, limited support for specific obfuscators, or poor scalability. For instance, the Java Deobfuscator repository has not received updates in over two years. For this task, Java Deobfuscator was tested because it is open source. However, testing revealed that Java Deobfuscator failed to detect obfuscation patterns in 30 malicious JAR files from the MalDICT dataset \cite{joyce2023maldictbenchmarkdatasetsmalware}. Due to the limited availability of open-source deobfuscators for Java, we created a custom deobfuscation tool using Java and the JavaParser library \cite{javaparser}. 
Our deobfuscator parses Java source code, walks its AST (Abstract Syntax Tree), evaluates constant-like expressions (especially string-related ones), and replaces obfuscated expressions in the source with their computed literal values. The approach effectively deobfuscates Java code by resolving and inlining constant string expressions. Although the approach does not address all obfuscation techniques and may not succeed in every case, it allows us to assess LLMs' ability to understand the code by potentially revealing obfuscated strings. Figure \ref{fig:code-diff-deobfuscated} provides an example and highlights the differences in the code before and after applying our deobfuscator. Red lines indicate the original code (decompiled but not yet deobfuscated), while green lines represent the code after applying deobfuscation. 

\subsubsection{Summarization of the code} \label{sec:method-summarisation}

The third step is to summarize the code using an LLM. This step answers  \textbf{RQ1} by including a verdict (i.e., classifying the file as Benign or Malicious, and providing a confidence score and rationale) in the summary. 
This step also contributes to answering \textbf{RQ2} by extracting knowledge and keywords that can be searched in the vector database, thereby identifying relevant CVEs. Summarizing the code is done by prompting the LLM with the decompiled and deobfuscated code, along with an instruction prompt. Figure \ref{fig:example-prompt-summarize-malware} presents an example input prompt. This input enables the extraction of technical details that can be matched with CVE descriptions. The LLM generates a structured response, defined by a JSON schema containing specific properties. The response includes a verdict, summary, observed activities, indicators of compromise (IoCs), utilized libraries, and relevant CVE search queries. An example output is shown in Figure \ref{example-output-summary}.
\begin{figure}[!h]
\begin{tcolorbox} [title=Prompt, colback=gray!20]
    Act as a malware analyst and thoroughly examine this decompiled executable code. Break it down step by step, identifying key behaviors, suspicious patterns, and potential vulnerabilities.\newline\newline
    1. \textbf{Verdict}: Classify the file as Benign, Suspicious, or Malicious, with a confidence score and rationale.\newline
    2. \textbf{Summary}: Provide a clear explanation of what this code does.\newline
    3. \textbf{Activities}: List all major operations performed (e.g., file manipulation, registry modification, network access).\newline
    4. \textbf{Indicators of Compromise (IOCs)}: Extract URLs, created files, registry changes, mutex usage, and network activity.\newline
    5. \textbf{Libraries Used}: Identify dependencies and third-party modules.\newline
    6. \textbf{Threat Mapping}: Map activities to known MITRE ATT\&CK tactics and techniques.\newline
    7. \textbf{CVE Search Queries}: Generate contextually enriched search queries for finding vulnerabilities.\newline
    
    Do not focus only on obfuscation or anti-analysis techniques for determining the nature of the code.\newline

    File 1:\newline
    \{code\}\newline

    File 2:\newline
    \{code\}
\end{tcolorbox}
    \caption{Example Prompt: Summarizing Decompiled Malware (To save space, the decompiled code is substituted with a placeholder.)}
    \label{fig:example-prompt-summarize-malware}
\end{figure}

\begin{figure}
 \centering

\begin{tcolorbox}[title=Generated Summary]
\begin{lstlisting}[language=json,firstnumber=1]
{
  "verdict": "Malicious",
  "summary": "The code is designed to bypass Java's built-in 
  security mechanisms, download an executable file
  from a specified URL, and execute it. 
  This is indicative of a trojan behavior.",
  "activities": [
    "Disabling Java Security Manager settings",
    "Downloading an executable file from a URL",
    "Executing a downloaded executable file",
    "File manipulation in the user's temporary directory"
  ],
  "indicators_of_compromise": {
    "urls": [
      "(url string from decoded sb2)"
    ],
    "created_files": [
      "(tmp file created using sortBy function)"
    ],
    "registry_entries": [],
    "mutex": [],
    "network_activity": [
      "Download of executable from URL"
    ]
  },
  "libraries_used": [
    "java.io.InputStream",
    "java.io.FileOutputStream",
    "java.security.AccessControlContext",
    "java.security.Permission",
    "java.security.AllPermission",
    "java.net.URL",
    "javax.applet.Applet"
  ],
  "cve_search_queries": {
    "status": "success",
    "message": "Generated CVE search queries for potential 
    vulnerabilities related to Java security bypass techniques.",
    "keywords": [
      "Java Applet security bypass CVEs",
      "Java deserialization vulnerability CVEs",
      "Java insecure download CVEs"
    ]
  }
}
\end{lstlisting}
\end{tcolorbox}
\caption{An Example Output}
\label{example-output-summary}
\end{figure}

\subsubsection{Generating Search Queries} \label{sec:method-generating-search-queries}

The fourth step involves issuing a separate prompt to generate additional search queries based on the summary. This approach allows the system to use a more powerful LLM while limiting the number of input tokens by excluding the code and incorporating only the summary generated in the previous step.
Figure \ref{fig:example-prompt-generate-search-queries} shows the prompt that is passed into the LLM. The prompt is designed to reduce the generation of vague queries that refer to obfuscated identifiers or overly generic descriptions. The generated search queries are returned in plain text format and forwarded to the next step of the analysis flow to answer \textbf{RQ2}.

\begin{figure}
\begin{tcolorbox} [title=Prompt, colback=gray!20]
    Act as a cybersecurity expert specializing in malware analysis. Given a decompiled Java executable summary, extract only the most relevant attack techniques, APIs, libraries, and execution patterns that could be linked to known CVEs. DO NOT include:\newline
    
    - Random class names (aoaz, ykybiswclildke).\newline
    - Generic words like temporary directory, specific integer, arbitrary parameters.\newline
    - Broad file names or placeholders (update.exe, mrun).\newline
    - Vague descriptions like 'functions invoking methods through reflection' - instead, specify the function names.\newline
    
    Focus Areas:\newline
    - APIs \& Methods: (e.g., getMethod(), invoke(), defineClass(), definePackage(), JMX)\newline
    - Exploit Techniques: (e.g., 'Remote class loading via ClassLoader', 'Bypassing security manager using reflection')\newline
    - Indicators of Compromise (IOCs): (e.g., suspicious URLs, registry changes, system modifications)\newline
    - Vulnerable Components: (e.g., 'JMX MBeans exposure', 'Unsigned Applet execution')\newline
    
    Limit the queries to the 5-10 most relevant terms that could be used to search for known vulnerabilities.\newline

    <generated summary>
\end{tcolorbox}
    \caption{Prompt: Generate Search Queries}
    \label{fig:example-prompt-generate-search-queries}
\end{figure}

\subsubsection{Semantic Similarity Search} \label{sec:method-similarity-search-details}

This step implements the retrieval-augmented generation (RAG) component of the system to address \textbf{RQ2}. Its function is to identify the most semantically relevant CVE descriptions for a given input malware by comparing their respective vector embeddings. The relevant CVEs are incorporated into the context of the LLM for use in Step 7, as detailed in Section \ref{sec:method-predict-cve-detailed}.

\begin{figure}
    \centering
    \includegraphics[width=1\linewidth]{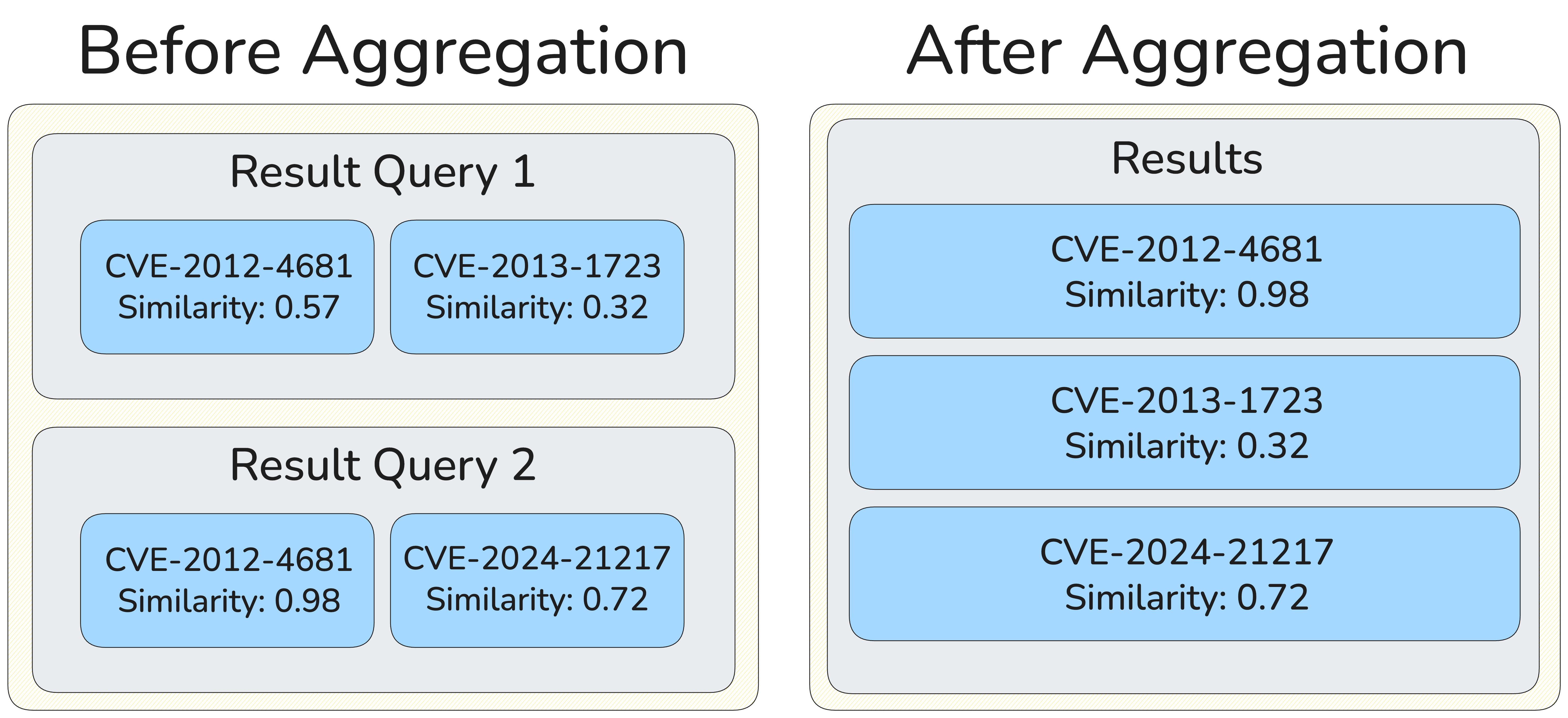}
    \caption{An Example of  Result Queries Before and After Aggregation}
    \label{fig:aggregation-example}
\end{figure}

After the code summary and search queries have been generated, as described in Sections \ref{sec:method-summarisation} and  \ref{sec:method-generating-search-queries}, the queries are transformed into a dense vector representation. Each query is mapped to a single vector, which is subsequently used to perform an \textbf{approximate nearest neighbor (ANN)} \cite{milvus-vector-search} search for each vector against a database (i.e., the Milvus database described in Section \ref{sec:cve-database-overview}). Each search retrieves up to 100 entries. This reduces the risk of overlooking a relevant CVE and increases the chance that the ranking step finds a better-fitting CVE based on exact keywords. Additionally, the results of each search are aggregated by selecting the maximum similarity score for each unique CVE. An example of the aggregation is shown in Figure \ref{fig:aggregation-example}. This ensures that a CVE is not duplicated in the final results.

\subsubsection{Ranking the Semantic Search Result} \label{sec:method-reranking-detailed}

In this step, MalCVE ranks the CVEs identified in the previous step using the BM25 algorithm \cite{okapiBM25}. BM25 was selected for its open-source nature and minimal computational requirements, including no GPU requirement. The similarity search in Step 5 captures high-level semantic similarities. BM25 measures direct lexical overlap between the analyzed code and CVE descriptions. This is particularly useful for identifying whether specific libraries are referenced in the CVE descriptions.

The libraries referenced in the generated summary are extracted for further analysis. Each library name is tokenized and compiled into a flat list. Generic tokens, such as Java, are excluded to minimize noise and potential bias. Next, the descriptions of each CVE in the top-k results from the semantic similarity search are tokenized. The tokenized library names are then used to query the CVE descriptions.

The original similarity scores and BM25 scores are normalized to ensure proportional contribution to the final score. The final score is computed as a \textbf{weighted combination} of the normalized similarity and BM25 scores, using the formula: 
Final Score = 0.7 x Normalized Similarity Score + 0.3 x Normalized BM25 Score.

When extracting CVE information, CWE identifiers associated with the CVEs are included when available. After ranking, CVEs sharing the same CWE are retrieved, expanding the candidate set to include vulnerabilities with the same root cause.

\subsubsection{Associating malicious JAR files to CVEs} \label{sec:method-predict-cve-detailed}

To answer \textbf{RQ2}, the system associates the malicious JAR files with relevant CVEs. Figure \ref{fig:example-prompt-predict-cve}  shows the prompt template for associating the relevant CVEs. The prompt includes the code summary, the top 10 CVEs retrieved from the similarity search, and the deobfuscated code to provide the LLM with sufficient contextual information for accurate association.

The LLM produces output in structured JSON format, with the schema defined in the request. This approach ensures consistency across associations. The response schema includes three properties:

\begin{enumerate}
    \item Behavior Explanation: A brief explanation of the behavior described in the summary.
    \item Matched CVE: The CVE number that best fits the behavior described in the summary.
    \item Justification: A short explanation of why the selected CVE matches the observed behavior.
\end{enumerate}

\begin{figure}
\begin{tcolorbox} [title=Prompt, colback=gray!20]
    \# Act as a cybersecurity expert specializing in malware analysis.\newline
    Based on the previous summary and the following CVEs, your task is to evaluate and determine which CVE from the list below that fits the behaviour described in the summary.\newline\newline
    \#\# Summary of decompiled jar file: \\
    <summary>\newline\newline
    \#\# List of CVEs to evaluate:\newline
    <cves>\newline~\newline
    Start with a brief explanation of the behaviour described in the summary, then list the CVE number that best fits the behaviour. Provide a short explanation of why you chose that CVE.\newline\newline
    \#\# Deobfuscated Code:\newline
    <code>
\end{tcolorbox}
    \caption{Prompt: Predict CVE}
    \label{fig:example-prompt-predict-cve}
\end{figure}

\subsubsection{Save Results to File}

The final step in the analysis pipeline involves storing the prediction results, the generated summary, the search queries, and metadata in a single file. The metadata includes information about the models and prompts utilized during the analysis.

\subsection{MalCVE implementation}

\begin{figure}
    \centering
    \includegraphics[width=1\linewidth]{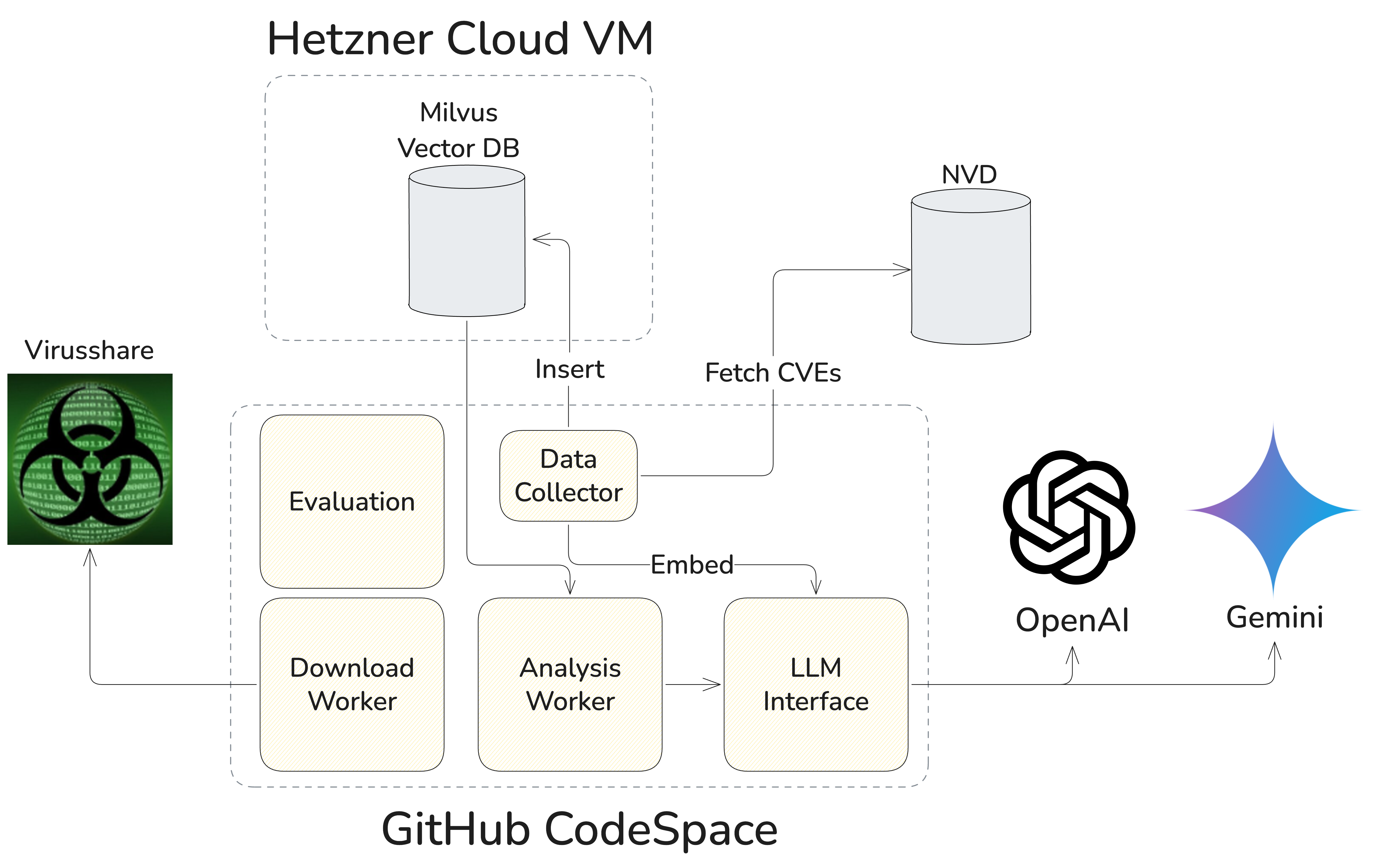}
    \caption{System Architecture Diagram}
    \label{fig:architecture-overview}
\end{figure}

Figure \ref{fig:architecture-overview} illustrates the architecture of the proof-of-concept tool MalCVE. The system includes a development and execution environment composed of GitHub Codespace and a Hetzner Cloud virtual machine (VM) that hosts the database.

\subsubsection{GitHub Codespace}

GitHub Codespaces are used to develop and execute a Python script that includes the Data Collector (Section \ref{sec:data-collector}), which initializes the Milvus Vector Database (Section \ref{sec:cve-database-overview}). The script also comprises the Download Worker (Section \ref{method:download-worker}), Analysis Worker (Section \ref{method:analysis-worker}), LLM Interface (Section \ref{sec:method-llm-interface}), and an Evaluation module (Section \ref{sec:method-evaluation}).

\subsubsection{Data Collector} \label{sec:data-collector}

The data collector script processes data from the National Vulnerability Database (NVD) and inserts it into the Milvus Vector Database (see Section \ref{sec:cve-database-overview}).
Prior to insertion, each CVE description is transformed into a 1536-dimensional vector using OpenAI’s text-embedding-3-small model. This model was selected for its computational efficiency and lower cost, which are advantageous given the large volume of CVE entries.

All available technical details from the National Vulnerability Database (NVD) were stored to enable advanced querying and filtering capabilities.
A notable enhancement was the addition of Common Weakness Enumeration (CWE) identifiers. By storing the CWEs associated with each CVE, the system can identify vulnerabilities that share a common root cause. This feature is particularly useful when analyzing CVEs with high similarity scores (see Section \ref{sec:method-reranking-detailed}).
Although the Common Vulnerability Scoring System (CVSS) base score and vector string are not currently used in the analysis, they were included to support future filtering by severity.
In the final system, the data collector stored the following data:

\begin{itemize}
    \item \textbf{CVE ID:} Unique identifier for the CVE.
    \item \textbf{Description:} Textual description of the CVE.
    \item \textbf{Description Vector:} Vectorized representation of the description.
    \item \textbf{CWE:} Common Weakness Enumerations related to the CVE.
    \item \textbf{CVSS:} The CVEs Common Vulnerability Scoring System string representation.
    \item \textbf{CVSS Score:} The CVEs CVSS base score.
\end{itemize}

\subsubsection{Hetzner VM and Vector Database} \label{sec:cve-database-overview}

A Hetzner Cloud VM hosts the Milvus vector database, which stores CVEs. Hosting the database on a separate VM isolates the database from GitHub Codespaces, where malicious files are handled. The database functions as a knowledge base by incorporating relevant CVEs into the context used for the final prompt in Step 7 (see Section \ref{sec:method-predict-cve-detailed}).
The system is configured to periodically fetch updates from NVD, keeping the database current. This is essential for an LLM to accurately identify newly published CVEs.

\subsubsection{Download Worker} \label{method:download-worker}

The system adds a list of file hashes to the download queue. The Download Worker retrieves and downloads the corresponding files from VirusShare, adhering to the platform’s rate limits. Successfully downloaded files are then transferred to the analysis queue for further processing.

\subsubsection{Analysis Worker} \label{method:analysis-worker}

The Analysis Worker is responsible for decompiling, deobfuscating, summarizing, generating search queries, performing semantic search, ranking results, generating CVE predictions, and saving the output to a file. Multiple Analysis Workers can operate concurrently. Each worker retrieves a file from the analysis queue and performs the full sequence of tasks to complete the analysis.

\subsubsection{LLM Interface} \label{sec:method-llm-interface}

All components interact with LLMs through a consistent interface that enables file summarization, search query generation, and final association output. This design facilitates the integration of additional LLMs in future implementations.

\section{Evaluation}
\label{sec:evaluation}
This section presents the design and implementation of the evaluation for MalCVE,  along with the corresponding results. 

\subsection{Evaluation Design}
\label{sec:method-evaluation}
To evaluate RQ1, a dataset containing malware samples is required. For RQ2, a dataset containing both malware samples and their corresponding CVE identifiers is required.

MalDICT \cite{joyce2023maldictbenchmarkdatasetsmalware} was selected as the benchmarking dataset. As noted by the authors, ``MalDICT includes the first public malware datasets labeled according to platform, vulnerability (CVEs), and packer'' \cite{joyce2023maldictbenchmarkdatasetsmalware}. The vulnerability subset of MalDICT contains 173,886 malware samples and annotations for 128 distinct vulnerability types \cite{MaldictGitHub:online}. From this dataset, we repeatedly selected decompilable files at random until the available tokens within our budget were insufficient for further analysis. Consequently, a total of 3,839 decompilable files were selected for analysis.

Our approach is not dependent on any specific LLM. To demonstrate its effectiveness, we conducted a comprehensive benchmark of several state-of-the-art LLMs. Due to practical constraints, testing was primarily limited to models from OpenAI and Gemini. Models provided by De because they failed to produce structured output in accordance with the required JSON schemas, hindering  result extraction and interpretation. Additional experimentation was limited by API restrictions, latency variability, and throughput constraints, particularly with OpenAI models, which were restricted to a rate of 200,000 tokens per minute.

For RQ1, we measured classification accuracy, i.e., the percentage of JARs that were correctly classified as either malicious or benign. For RQ2, we used recall@k where (k $\in$ {1, 3, 5, 10}) and accuracy as evaluation metrics. Recall@k assesses the effectiveness of the semantic similarity search by determining whether a correct CVE appears among the top-k MalCVE retrieved results. Accuracy, on the otmeasures the LLMs' ability to select the correct CVE. These metrics are also commonly reported in related work \cite{LLM-CloudSec}, \cite{vulrag}.

\begin{itemize}
    \item \textbf{Accuracy:} \(\frac{\text{\# Correct CVE Predictions}}{\text{\# Total CVE Predictions}}\)
    \item \textbf{Recall@k:} \(\frac{\text{\# Malicious samples with correct CVE in Top k}}{\text{\# Total Malicious Samples}}\)
 \end{itemize}

One challenge in associating malware with CVEs is that many CVEs lack sufficient technical detail. For example, CVE-2012-1723, which includes only version numbers and omits method or class names, is shown in Figure \ref{fig:example-cve-noinfo}. In contrast, CVE-2012-4681 provides more comprehensive technical information and is illustrated in Figure \ref{fig:example-cve-withinfo}.

\begin{figure}
\begin{tcolorbox} [title=CVE-2012-1723, colback=gray!20]
"Unspecified vulnerability in the Java Runtime Environment (JRE) component in Oracle Java SE 7 update 4 and earlier, 6 update 32 and earlier, 5 update 35 and earlier, and 1.4.2\_37 and earlier allows remote attackers to affect confidentiality, integrity, and availability via unknown vectors related to Hotspot" \cite{cve-unspecified}.
\end{tcolorbox}
    \caption{An example CVE without sufficient technical information}
    \label{fig:example-cve-noinfo}
\end{figure}

\begin{figure}
\begin{tcolorbox} [title=CVE-2012-4681, colback=gray!20]
"Multiple vulnerabilities in the Java Runtime Environment (JRE) component in Oracle Java SE 7 Update 6 and earlier allow remote attackers to execute arbitrary code via a crafted applet that bypasses SecurityManager restrictions by (1) using com.sun.beans.finder.ClassFinder.findClass and leveraging an exception with the forName method to access restricted classes from arbitrary packages such as sun.awt.SunToolkit, then (2) using "reflection with a trusted immediate caller" to leverage the getField method to access and modify private fields, as exploited in the wild in August 2012 using Gondzz.class and Gondvv.class" \cite{cve-2012-4681}.
\end{tcolorbox}
    \caption{An example CVE with sufficient technical information}
    \label{fig:example-cve-withinfo}
\end{figure}

Thus, we filtered the MalDICT dataset to include only CVEs with sufficient technical detail, specifically focusing on Java JAR files labeled with three CVE identifiers: CVE-2012-4681, CVE-2013-0422, and CVE-2013-1493. The decision to limit the scope to these three CVEs, rather than including more technically detailed ones, was influenced by budgetary constraints associated with conducting evaluations using commercial LLMs.

\subsection{Evaluation Results}
\subsubsection{RQ1: Malware Detection using LLMs}\label{sec:results:verdict}

Figure 10 presents the results of single-class malware detection, illustrating how effectively the LLM classifies binaries as either malicious or benign. To assess the impact of LLM architectural variations, multiple iterations were conducted. Each colored dot represents a model, and the number displayed on the dot indicates the number of files analyzed during that iteration.

\begin{figure}[!h]
  \centering
    \includegraphics[width=0.5\textwidth]{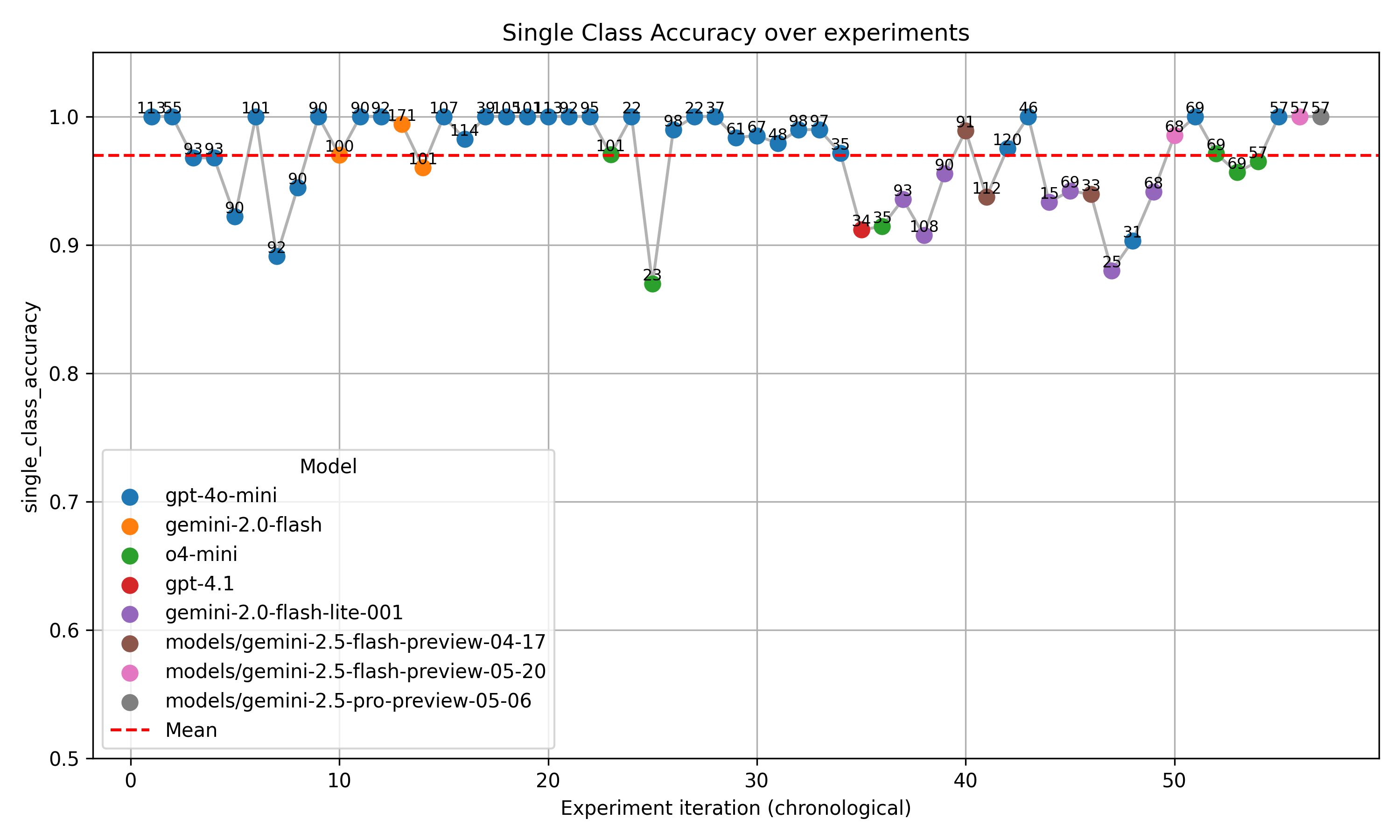}
  \caption{Results of RQ1}
  \label{fig:verdict-accuracy-trend}
\end{figure}

The dashed red lines in Figure \ref{fig:verdict-accuracy-trend} show that the mean accuracy is \textbf{97\%}, across all 3,839 analyzed JARs and LLMs. This corresponds to 115 misclassified files out of the total. The observed accuracy of different LLM models ranged from 88\% to 100\%.

The findings demonstrate that LLMs can correctly classify decompiled binaries as either malicious or benign uswith appropriate prompting. This was achieved without any fine-tuning, relying solely on zero-shot prompting. Changes in LLM architecture did not have a significant impact on accuracy, highlighting that LLMs erform effectively with default configurations for this task.

Throughout the evaluation, two commercial LLM providers were utilized: Google (Gemini models) and OpenAI.

\begin{itemize}
    \item \textbf{Google} incurred a cost of \$69.86 for completions using Gemini models
    \item \textbf{OpenAI} incurred a total cost of \$44.46, which includes embedding approximately 21.5 million tokens for semantic similarity search and processing completion requests totaling 68.5 million tokens.
\end{itemize}

In total, 3,839 JAR executables were analyzed during the evaluation. This corresponds to an average cost of approximately \textbf{\$0.03 per file}, inclusive of all prompting and embedding expenses. This per-file cost is roughly \textbf{1/66th} the cost of CrowdStrike Falcon’s \$2 per file and \textbf{1/80th} the cost of ANY.RUN’s \$2.40 per file.

\vspace{1em}
\noindent\fbox{%
    \parbox{\columnwidth}{%
        \textbf{Summary for RQ1}: MalCVE using LLMs combined with decompilation and deobfuscation can accurately identify malware in a cost-effective manner.
    }%
}
\hfill \break
\subsubsection{RQ2: CVE association} 
\label{sec:rq2-results}

To answer RQ2, we analyzed the performance of various LLMs to identify the best-performing systems. The number of files analyzed per model varied due to budgetary constraints, limiting a full evaluation of all models across the entire dataset. In total, 620 JAR files were tested, distributed among the models as follows: O4-Mini (195 files), GPT-4o Mini (157 files), Gemini 2.5 Flash (125 files), Gemini 2.5 Flash Lite (86 files), and Gemini 2.5 Flash Pro (57 files). Table \ref{table:final-system-metrics} presents classification accuracy and recall@k metrics, including their mean, maximum, minimum, and standard deviation values. 

Accuracy peaked at 44\%, with a mean of 39\%. Gemini 2.5 Pro achieved the highest accuracy, followed closely by Gemini 2.5 Flash and O4-Mini, both at approximately 42\%. GPT-4o Mini recorded the lowest accuracy at 26\%.

Mean values for Recall@1, Recall@3, and Recall@5 were 22\%, 35\%, and 42\%, respectively. Gemini 2.5 Pro consistently scored the highest across these metrics, with O4-Mini ranking second. The highest observed Recall@10 was 65\% (O4-Mini), with a mean of 53\%. While the average Recall@10 is moderate, the maximum value suggests that semantic similarity search can be further optimized to achieve satisfactory performance.

\begin{table}
\centering
\caption{Results of RQ2 (Files Analyzed: 620)}
\label{table:final-system-metrics}
\begin{tabular}{lccccc}
\hline
\textbf{Metric} & \textbf{Mean} & \textbf{Max} & \textbf{Min} & \textbf{Std Dev} \\
\hline
\textbf{Accuracy}  & 0.39 & 0.44 & 0.26 & 0.06 \\
\textbf{Recall@1}              & 0.22 & 0.32 & 0.12 & 0.06 \\
\textbf{Recall@3}              & 0.35 & 0.46 & 0.22 & 0.08 \\
\textbf{Recall@5}              & 0.42 & 0.51 & 0.28 & 0.08 \\
\textbf{Recall@10}             & 0.53 & 0.65 & 0.40 & 0.08 \\
\hline
\end{tabular}
\end{table}

\textbf{Impact of technical informtion in CVE descriptions:} As explained in Section \ref{sec:evaluation}, LLMs may fail to associate malware with specific CVEs when the CVE descriptions lack sufficient technical detail. Therefore, our evaluation focused exclusively on CVEs that provide adequate technical information.

To assess the impact of technical information in CVE descriptions, we included additional CVEs lacking sufficient detail in the analysis to evaluate performance differences. The results show that Recall@10 decreased by approximately 8\%, Recall@5 by 11\%, and Recall@3 by 6\%, while Recall@1 remained largely unchanged. Accuracy also declined by 8\%. These findings underscore the importance of detailed CVE descriptions for extracting relevant vulnerabilities, as LLMs are unable to reliably associate malware with CVEs when descriptions lack technical detail beyond the affected Java version. This highlights the value of publicizing and sharing comprehensive CVE information. However, expecting every CVE to include detailed technical descriptions may be impractical. Therefore, incorporating data from additional sources, such as CWEs, or crawling advisory references published by the NVD, may help improve the accuracy of associating the malware with CVEs.

\textbf{Impact of deobfuscation:} Existing studies \cite{walton2025exploringlargelanguagemodels}, \cite{PE-Ransomware-Image-Detection} do not account for deobfuscation. As part of a supplementary analysis, we removed the deobfuscation step in MalCVE to evaluate its impact on performance. Following this removal, Recall@10 decreased by 13\%, Recall@5 by 10\%, Recall@3 by 9\%, and Recall@1 by 6\%. Accuracy also declined by 8\%. This decline is attributed to the LLM’s reduced ability to extract specific libraries and methods when deobfuscation is not applied. For example, the deobfuscator recovered critical strings indicating a call to "findClass" that were hidden in the decompiled code. This is important because CVE-2012-4681 is defined in terms of the misuse of "ClassFinder.findClass" and related methods. If those identifiers remain obfuscated, the model can’t connect the code to the vulnerability. The search queries used in semantic similarity search are critical for retrieving relevant CVEs.
Prior to deobfuscation, the extracted queries were relatively generic and frequently repeated. After applying deobfuscation, the queries became more diverse and detailed, often including specific libraries, classes, and methods (e.g., sun.awt.SunToolkit.getField, java.beans.Statement.setSecurityManager).

\vspace{1em}
\noindent\fbox{%
    \parbox{\columnwidth}{%
        \textbf{Summary for RQ2}: MalCVE can identify malware-associated CVEs with moderate of accuracy. The absence of sufficient technical detail in CVE descriptions presents a significant barrier to achieving higher accuracy.
    }%
}
\hfill \break

\section{Discussion}
\label{sec:discussion}

In this section, we discuss our approach and evaluation results, compare them with related work, and illustrate their implications for academia and industry. We also examine potential threats to the validity of our findings and describe the measures taken to address them. 

\subsection{Comparison with related work}

\citet{malbert} reported a 97.6\% accuracy in malware detection using a fine-tuned model. In comparison, our approach achieved a mean accuracy of 97\%, with some models reaching 100\%. These results provide evidence that generic LLMs, when combined with effective decompilation and deobfuscation techniques, can achieve malware detection performance comparable to that of fine-tuned models.

Previous LLM-based malware detection studies generally operated on fewer than \emph{250} samples. Walton et al. \cite{walton2025exploringlargelanguagemodels} reported results on only 200 Android APKs. The PE ransomware workwork of \cite{PE-Ransomware-Image-Detection} used an imbalanced set of 221 executables. Our evaluation covers 3,839 JAR files, yielding a much stronger statistical confidence. Additionally, we compared multiple LLM models and showed the generalizability of our approach. 

Both \citet{google-malware-analysis} and \citet{walton2025exploringlargelanguagemodels} rely on prompt engineering approaches that require a security expert to manually inspect and refine the model’s narrative outputs. In contrast, our pipeline is fully automated. A fixed zero-shot prompt generates a structured JSON verdict that can be directly integrated into downstream systems. This approach reduces, and in some cases eliminates, the need for human intervention in determining whether a file is malicious or benign.

\citet{walton2025exploringlargelanguagemodels} explicitly excluded obfuscated APKs and, as a result, did not report data on them. In contrast, our system decompiled and applied light deobfuscation to JAR files prior to classification, achieving a mean single-class accuracy of 97\% and a minimum observed accuracy of 88\%, as shown in Figure \ref{fig:verdict-accuracy-trend}. This level of performance is comparable to that of commercially available malware detection tools \footnote{https://www.av-comparatives.org/comparison/}.

We also investigated the ability of LLMs to associate specific CVEs with decompiled malicious binaries using prompting, light deobfuscation, and RAG. Existing studies \cite{LLM-CloudSec}, \cite{vulrag} focus primarily on CWE classification in source code, which represents a substantially more general task compared to CVE-level association. The presence of obfuscated binaries and the extensive CVE vocabulary, exceeding 280,000 entries, makes CVE-level association significantly more challenging. Despite these difficulties, our results match or exceed those reported in source-code-based studies on several metrics. For instance, our highest Recall@10 was 65\%, outperforming the 61\% Recall@10-30 reported by \citet{vulrag}.

\subsection{Implications for Industry and Academia} \label{sec:implications}

Achieving 97\% accuracy without fine-tuning demonstrates that generic LLMs are already viable as lightweight, scalable malware filters in real-world security workflows. At a cost of \$0.03 per file, MalCVE is significantly more cost-effective than commercial alternatives, such as CrowdStrike Falcon and ANY.RUN. In the MalCVE implementation, all components, except the commercial LLMs, are open source, enabling users to replicate or customize the solution without incurring licensing fees. Security vendors can integrate the solution into existing products with minimal overhead. In particular, vendors offering analysis tools \textbf{can leverage LLM-generated summaries to provide explainable verdicts to customers who require insight into why a file was flagged as malicious.} Notably, this also enables small organizations to deploy a malware filter without the upfront costs associated with commercial tools.

However, a significant limitation arises from the LLM context window constraints. OpenAI models, including GPT-4o-mini, cannot accommodate large JAR binaries due to limited token capacity. Even models with extensive context windows, such as Gemini 1.5 and 2.0 (1,000,000 tokens), occasionally struggled with comprehensive semantic analysis of large benign binaries. Consequently, complex files may be inadequately summarized or misclassified due to truncated input, reducing the construct validity of the semantic representation approach, although newer models might not suffer from the same issues this study encountered.

Our findings for RQ2 indicate that all evaluated models show potential for CVE association. This means that LLMs can support rapid analysis of suspicious files, enabling the identification of relevant CVEs and informing potential mitigation strategies. Furthermore, organizations can select models based on performance and cost considerations, making the solution accessible even to smaller organizations with limited resources. Although the current implementation focuses on Java, the approach is generalizable to other programming languages by substituting the decompiler and deobfuscator with language-specific tools.

From an academic standpoint, this work introduces a promising research direction in CVE-level association for malicious binaries using LLMs. As the first study to apply LLMs to CVE association in decompiled malware binaries, it establishes a foundational benchmark for future research.

\subsection{Threats to Validity}

\subsubsection{Internal Validity}
LLM performance is strongly influenced by prompt formulation. Small variations in prompting can affect model outcomes. Although zero-shot prompting was consistently applied, subtle modifications could still introduce bias and influence model predictions. However, our study shows that all evaluated LLMs demonstrated satisfactory performance in detecting malware, indicating that the impact of slight prompt variations may be negligible. 

\subsubsection{External Validity}
The decompilation and deobfuscation processes in MalCVE are specific to Java. Consequently, the LLM's ability to interpret the code depends on the quality of these preprocessing steps. Although the dataset was substantial in size, it may not adequately represent the diversity of real-world malware. This limitation affects the generalizability of the findings to the full spectrum of potential malware behaviors. Future work will involve conducting further research on malware written in other programming languages and bigger datasets. 

Open-source LLMs are also viable options for integration into MalCVE to detect malware. However, this study focuses primarily on commercial models, as many companies may be reluctant to input their code into open-source LLMs due to security concerns. Focusing on commercial models also enables a comparative analysis of the costs associated with using MalCVE versus commercial malware detectors. A future study is planned to evaluate the performance of open-source LLMs within the MalCVE framework.

\subsubsection{Construct Validity}

Filtering MalDICT to retain only vulnerabilities with detailed technical descriptions improved accuracy by reducing semantic ambiguity, but also reduced the dataset's diversity. However, mapping malware to CVEs may not be feasible without technical information.

Malware might exploit multiple CVEs. The MalDICT dataset and our evaluation employed a singused single-label classification, labeling correct but secondary CVEs as false negatives. This methodological limitation artificially reduces precision and may lead to an underestimation of MalCVE's performance. However, no other available datasets provide more precise associations between malware and CVEs. 

A common issue in evaluating LLM-based applications is data leakage. Since access to the MalDICT dataset requires authentication and the data consist of malware in binary format, it is unlikely this dataset was used to train commercial LLMs. Therefore, we believe the risk of data leakage in our evaluation is minimal. 

\subsubsection{Reliability}


The experiments were restricted by limited resources, totaling approximately \$58.75 for OpenAI services and \$69.86 for Gemini across 3,839 analyzed JARs. We did not compare the accuracy of MalCVE with commercial tools due to their high cost. Our limited budget constrained the scale and depth of the experiments relative to typical commercial research standards. Larger and more diverse experiments, which are often feasible in industry-funded studies, were infeasible within the budget and scope of our study.



\section{Conclusion and future work}
\label{sec:conclusion}

Malware can cause significant societal consequences. This study investigated the use of decompilation and deobfuscation paired with LLMs to detect whether binary files are malicious. The approach involved decompiling and deobfuscating the code, after which a LLM summarized the program’s behavior and rendered a verdict. The solution achieved a 97\% mean accuracy across 3,839 analyzed JAR binaries, offering a low-cost, low-effort alternative to dynamic analysis tools, fine-tuning approaches, and commercial platforms. We also investigated the use of LLMs and RAG to associate malware with the CVEs it may potentially exploit. The results demonstrate that LLMs supported by deobfuscation and semantic similarity search can satisfactorily associate malware with CVEs, although there remains room for improvement in accuracy.

Future work will aim to extend this study to support multiple programming languages, integrate open-source LLMs, and incorporate additional techniques, such as YARA rules, to improve accuracy in associating malware with CVEs.

\section{Data availability}
\noindent The software and data replication packages are published at:\\
\url{https://doi.org/10.5281/zenodo.16233415}\\

\bibliographystyle{ACM-Reference-Format}
\bibliography{reference}


\end{document}